# Micro force Measurement by an Optical Method

*Naceur-Eddine Khélifa*

LNE-INM / Conservatoire national des arts et métiers
61 rue du Landy, 93210 la Plaine Saint Denis, France

**Abstract** − We discuss the application of stress-induced changes in the crystal of a monolithic Nd:YAG laser as a possibility for microforce measurement. In fact, application of an unknown force on the resonator-amplifier-crystal can lead to a several gigahertz change, depending on the force intensity, of the laser frequency. In addition, the changes rates of the two orthogonally polarizations of the same mode with applied force are different. Hence, the strength of the applied stress can be deduced from measurement of induced change in the beat frequency between the two polarizations or between the fast mode (mode polarized in the orthogonal direction of the stress) and a reference frequency.

## 1. INTRODUCTION

The measurement of force at the micro or nano newton level is becoming essential for many sectors of science, industry and major balance companies. In fact, the traceability of the smallest forces has become a need of nanometrology because a variety of activities are generated in nanotechnology and biotechnology. We note some interesting studies on the electrostatic force traceable to the SI unit [1] [2] [3] but also on force generated by mass standards, which can be now extended to about $1\mu N$ [4].

The challenge of our project is to use photo-elastic effect in a solid-state laser [5][6] to measure small forces. Generally, elastic stress cause changes of the refractive index proportional to the applied force. This index change (by induced birefringence) is different for the electric vector for the light wave parallel and normal to the direction of the applied uniaxial stress.

The consequence of the stress-induced birefringence is that the monolithic (resonator-amplifier) Nd:YAG laser can be operated simultaneously in two orthogonally polarized modes. These modes have the particularity of exhibiting different tuning rates with applied stress. In practice, the stress is applied transversely to the rod axis of the monolithic resonator. In this case, the cavity mode polarized perpendicular to the applied stress called "the fast mode" tunes much more rapidly than "the slow mode" polarized parallel to the stress. The tuning rate of the fast mode is about 350 MHz/N [7].

## 2. PRINCIPLE OF MEASUREMENT

The induced birefringence splits the polarization degeneracy, leading to two orthogonal polarizations for the laser cavity. Here, we consider the beat frequency $\Delta \nu_i = \nu_{q,x} - \nu_{q,y}$ between the two frequencies having the same order $q$. For an elastic, mechanically homogeneous laser crystal, the birefringence induced by a transverse force F, leads to a change of the beat frequency between the two polarized modes given by:

$$\Delta \nu_i = K_a \frac{C \times F}{\lambda \times \ell \times d}$$

Where $\ell$ is the length of the crystal and $d$ its diameter, $\lambda$ the laser wavelength and $C$ is the optical stress coefficient that characterizes the photo elastic crystal [expressed in Brewster = $10^{-12}$ m$^2$/N]; $K_a$ is a coefficient, which depends on the shape of the crystal and the geometrical alignment.

By using a sufficiently small resonator structure, on can obtain single longitudinal mode oscillation. This configuration is obtained with the resonator mirrors coated upon the Nd:YAG crystal end faces. The advantage of this structure of resonator is that laser emission is in a way immune to misalignment from external noise.
Owing to the induced birefringence by the external force, each cavity mode splits into two orthogonally polarized resonances $\nu_{q,x}$ and $\nu_{q,y}$ as indicated by figure 1.

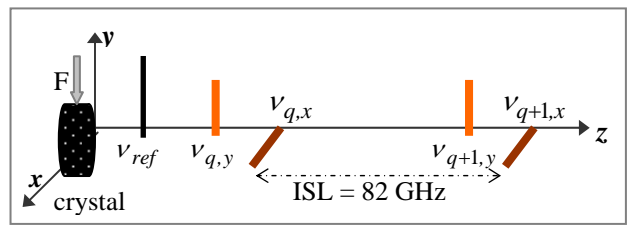

Fig. 1. Wavelength of fast and slow modes of a laser operated in a longitudinal monomode emission.

In practice, the beat frequencies $\Delta \nu_{ir} = \nu_{ref} - \nu_{q,x}$ and also $\Delta \nu_i = \nu_{q,x} - \nu_{q,y}$ can be measured easily. From these results, we hope to make:
- Static force calibration from a mass standard, if acceleration is well defined by using $F = M \times g$.
- Dynamic force calibration of micro probes (for example probing forces exerted by the cantilever of an atomic force microscope).

## 3. EXPERIMENT

The schematic diagram of the proposed optical micro-force sensor is similar to that previously reported [5] which uses laser emission at 1064 nm in Nd:YAG crystal. Our choice is to use crystal with small dimensions and to work at shorter wavelength emission [8] of the Nd:YAG laser ($\lambda = 946\, nm$). The system consists of a near infrared diode laser for optical pumping, a Nd:YAG mirrored crystal and a beat frequency measurement bench. A high doping density of the YAG matrix is considered for good pump absorption in a short length. The pump source consists of a 100 mW monomode diode laser. To minimize optical feedback, laser emission is collimated by a short- focal objective and sending through an optical isolator. Then the beam is focused into the monolithic cavity.

The Nd:YAG rod, we plan to use, will play the role both of monolithic laser amplification and resonator. Two rods will be used which have a length of $\ell = 1$ and 1,5 mm and the same diameter of 2 mm. The pump input face of the resonator is polished to a few centimeter radius of curvature and coated for HT at 808,6 nm and HR at 946 nm. The output end face is plane and coated for 0,5 % transmission at 946 nm. The temperature of the laser diode pump is controlled and fixed so that the output wavelength is around 808,6 nm, which corresponds to the maximum of absorption band of Nd:YAG.

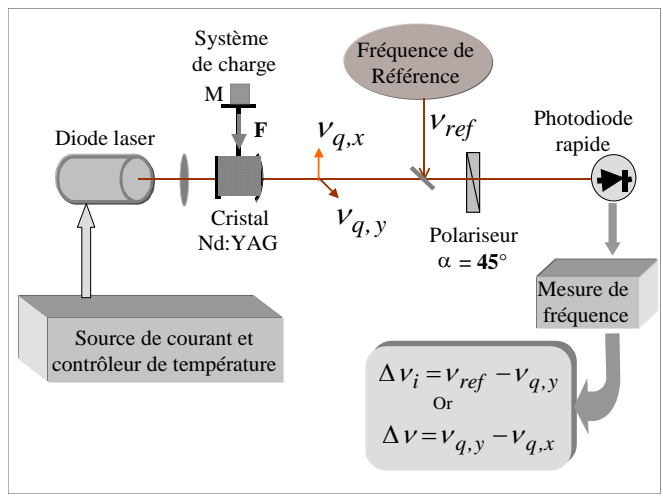

Fig. 2. Diagram illustrating the experimental set up.

## 4. CONCLUSION

An experimental investigation is under way at our institute. Improvement of resolution needs to get better stability of the beat frequency signal. Of course, this could be lead to a lower limit of force measurement. The level of $10^{-8}$ N is our realistic objective in this experiment. According to the theoretical previsions, the experiment can be used to connect masses standards in the range of $1\, kg$ to less than $100\, \mu g$.

TABLE 1. Approximate value of the beat frequency between the two polarizations for two crystal size and for $\lambda$ = 946 nm.

| Crystal dimensions (*mm*) | $\Delta \nu_i$ (*Hz*) for F = $10^{-7}$ N |
|---|---|
| $\ell$ = 1,5; D = 2, 0 | $\Delta \nu_i \approx 10$ |
| $\ell$ = 1,0; D = 2,0 | $\Delta \nu_i \approx 20$ |